\documentclass[aps,prx,amsmath,amssymb,reprint,superscriptaddress]{revtex4-1}
\usepackage{graphicx}
\usepackage{amsmath}
\usepackage{dcolumn}
\usepackage{bm}
\usepackage{bbold}
\usepackage{color}
\usepackage{amsfonts}
\usepackage{amssymb}
\usepackage{mathrsfs}
\usepackage{tabularx}
\usepackage{braket}
\usepackage{mathtools}
\usepackage{soul}			

\usepackage{caption}
\usepackage{subcaption}

\begin{document}

\title{A new method of probing mechanical losses of coatings at cryogenic temperatures}

\author{Serge Galliou}
\email{serge.galliou@femto-st.fr}
\affiliation{Department of Time and Frequency, FEMTO-ST Institute, UMR 6174, CNRS, UBFC, 26 rue de l'\'{E}pitaphe, 25030 Besan\c{c}on, France}

\author{Samuel Del\'{e}glise}
\email{deleglise@lkb.upmc.fr}
\affiliation{Laboratoire Kastler Brossel, UPMC-Sorbonne Universit\'{e}s, CNRS, ENS-PSL Research University, Coll\`{e}ge de France,75005 Paris, France}

\author{Maxim Goryachev}
\email{maxim.goryachev@uwa.edu.au}
\affiliation{ARC Centre of Excellence for Engineered Quantum Systems, University of Western Australia, 35 Stirling Highway, Crawley WA 6009, Australia}

\author{Leonhard Neuhaus}
\affiliation{Laboratoire Kastler Brossel, UPMC-Sorbonne Universit\'{e}s, CNRS, ENS-PSL Research University, Coll\`{e}ge de France,75005 Paris, France}

\author{Gianpietro Cagnoli}
\email{g.cagnoli@lma.in2p3.fr}
\affiliation{Laboratoire des Mat\'{e}riaux Avanc\'{e}s, CNRS/IN2P3, 69622 Villeurbanne, France}
\affiliation{Universit\'{e} Claude Bernard Lyon I, 69622 Villeurbanne, Fance}

\author{Salim Zerkani}
\affiliation{Laboratoire Kastler Brossel, UPMC-Sorbonne Universit\'{e}s, CNRS, ENS-PSL Research University, Coll\`{e}ge de France,75005 Paris, France}

\author{Vincent Dolique}
\affiliation{Laboratoire des Mat\'{e}riaux Avanc\'{e}s, CNRS/IN2P3, 69622 Villeurbanne, France}

\author{Xavier Vacheret}
\affiliation{Department of Time and Frequency, FEMTO-ST Institute, UMR 6174, CNRS, UBFC, 26 rue de l'\'{E}pitaphe, 25030 Besan\c{c}on, France}

\author{Philippe Abb\'{e}}
\affiliation{Department of Time and Frequency, FEMTO-ST Institute, UMR 6174, CNRS, UBFC, 26 rue de l'\'{E}pitaphe, 25030 Besan\c{c}on, France}

\author{Laurent Pinard}
\affiliation{Laboratoire des Mat\'{e}riaux Avanc\'{e}s, CNRS/IN2P3, 69622 Villeurbanne, France}

\author{Christophe Michel}
\affiliation{Laboratoire des Mat\'{e}riaux Avanc\'{e}s, CNRS/IN2P3, 69622 Villeurbanne, France}

\author{Thibaut Karassouloff}
\affiliation{Laboratoire Kastler Brossel, UPMC-Sorbonne Universit\'{e}s, CNRS, ENS-PSL Research University, Coll\`{e}ge de France,75005 Paris, France}

\author{Tristan Briant}
\affiliation{Laboratoire Kastler Brossel, UPMC-Sorbonne Universit\'{e}s, CNRS, ENS-PSL Research University, Coll\`{e}ge de France,75005 Paris, France}

\author{Pierre-Fran\c{c}ois Cohadon}
\affiliation{Laboratoire Kastler Brossel, UPMC-Sorbonne Universit\'{e}s, CNRS, ENS-PSL Research University, Coll\`{e}ge de France,75005 Paris, France}

\author{Antoine Heidmann}
\affiliation{Laboratoire Kastler Brossel, UPMC-Sorbonne Universit\'{e}s, CNRS, ENS-PSL Research University, Coll\`{e}ge de France,75005 Paris, France}

\author{Michael E. Tobar}
\affiliation{ARC Centre of Excellence for Engineered Quantum Systems, University of Western Australia, 35 Stirling Highway, Crawley WA 6009, Australia}

\author{Roger Bourquin}
\affiliation{Department of Time and Frequency, FEMTO-ST Institute, UMR 6174, CNRS, UBFC, 26 rue de l'\'{E}pitaphe, 25030 Besan\c{c}on, France}

\date{\today}


\begin{abstract}

A new method of probing mechanical losses and comparing the corresponding deposition processes of metallic and dielectric coatings in 1-100 MHz frequency range and cryogenic temperatures is presented. The method is based on the use of extremely high-quality quartz acoustic cavities whose internal losses are orders of magnitude lower than any available coatings nowadays. The approach is demonstrated for
Chromium, Chromium/Gold and a multilayer tantala/silica coatings. The ${\rm Ta}_2{\rm O}_5/{\rm Si}{\rm O}_2$ coating has been found to exhibit a loss angle lower than $1.6\times10^{-5}$ near 30 {\rm MHz} at 4 {\rm K}. The results are compared to the previous measurements. 


\end{abstract}

\maketitle

\section{Introduction}

Bulk acoustic wave (BAW) resonators are used in a variety of fields such as ultra-stable frequency reference\cite{Riehle2004}, spectral filtering in analog circuits\cite{Catherinot2011}, or mass sensing\cite{Janshoff2000}. With the recent development in cryogenic quartz technology with its exceptional quality factors\cite{Aspelmeyer2014,Goryachev:2013aa}, new application areas have emerged\cite{galliouScRip2013}. In particular, BAW acoustic devices are proposed as a platform to probe the Lorentz Invariance in the matter sector\cite{Lo:2016aa}, high frequency gravity wave detection\cite{Goryachev:2014aa}, links between gravity and quantum mechanics\cite{Goryachev:2014ab}, quantum information manipulation\cite{Woolley2016}, and it might be sensitive to particular types of dark matter\cite{Arvanitaki:2016aa}. 
In this work, we propose a new application of the quartz BAW technology to measurements of losses in coatings of crystalline solids and estimation of quality of corresponding deposition processes. 

A quartz substrate is geometrically optimized to oscillate at a specific resonance frequency, and piezoelectricity is used to drive and monitor the motion of a trapped acoustic wave in the resonator. In the traditional applications, metal coating is the preferred method to couple to mechanical motion by depositing a pair of electrodes directly on the surface of the resonator. It is well known that the resonance frequency is mass-sensitive and makes such a resonator, or in other words, acoustic cavity\cite{galliouScRip2013}, usable as a sensor, a so-called quartz-crystal micro-balance. Moreover, a specific reactive layer can be deposited on the metallic coating to selectively detect a given component by mass-loading. Finally, the quality factor sensitivity can also be used to probe defects, additive components or contamination \cite{Rodahl1995,Dixon2008} where it is assumed to be the dominant effect of the measurement. The same assumption is made in the present work where due to extremely low losses of the crystal substrate, a high-quality quartz trapped-energy resonator, the overall system dissipation measured via its mechanical Quality factor is determined by coatings. With BAW quartz resonator losses decreasing from $10^{-7}$ to almost as low as $10^{-10}$ at liquid helium temperatures\cite{galliouScRip2013,Goryachev:2013aa}, the approach might be used applied for tests from $20$K and below. The numerical values can be obtained for modes spanning over almost two decades, 1-100MHz, as measurements of high overtone (OT) modes has been demonstrated\cite{Goryachev:2013aa}. 


In this paper we analyze the coating effects of two metals, Chromium/Gold and Chromium, and of the most usual Bragg-reflector, a tantala-silica multilayer. The metallic coatings are typical solutions for devices used for stable freuqncy references. The mechanical behavior of dielectric thin-film materials at low temperatures is of prime interest in the context of third-generation gravitational wave interferometers, where high reflectivity-coatings ($> 99.99\%$) with low-mechanical loss angle at low temperature constitute a critical requirement. Note that metallic coatings have also been considered to deliberately lower mechanical Quality factors in order to prevent 3-mode parametric instability in large-scale gravitational-wave interferometers~\cite{Gras2008}. 

\section{Experimental technique}
\begin{figure}[t!]
\centering
\includegraphics[width=3.25in]{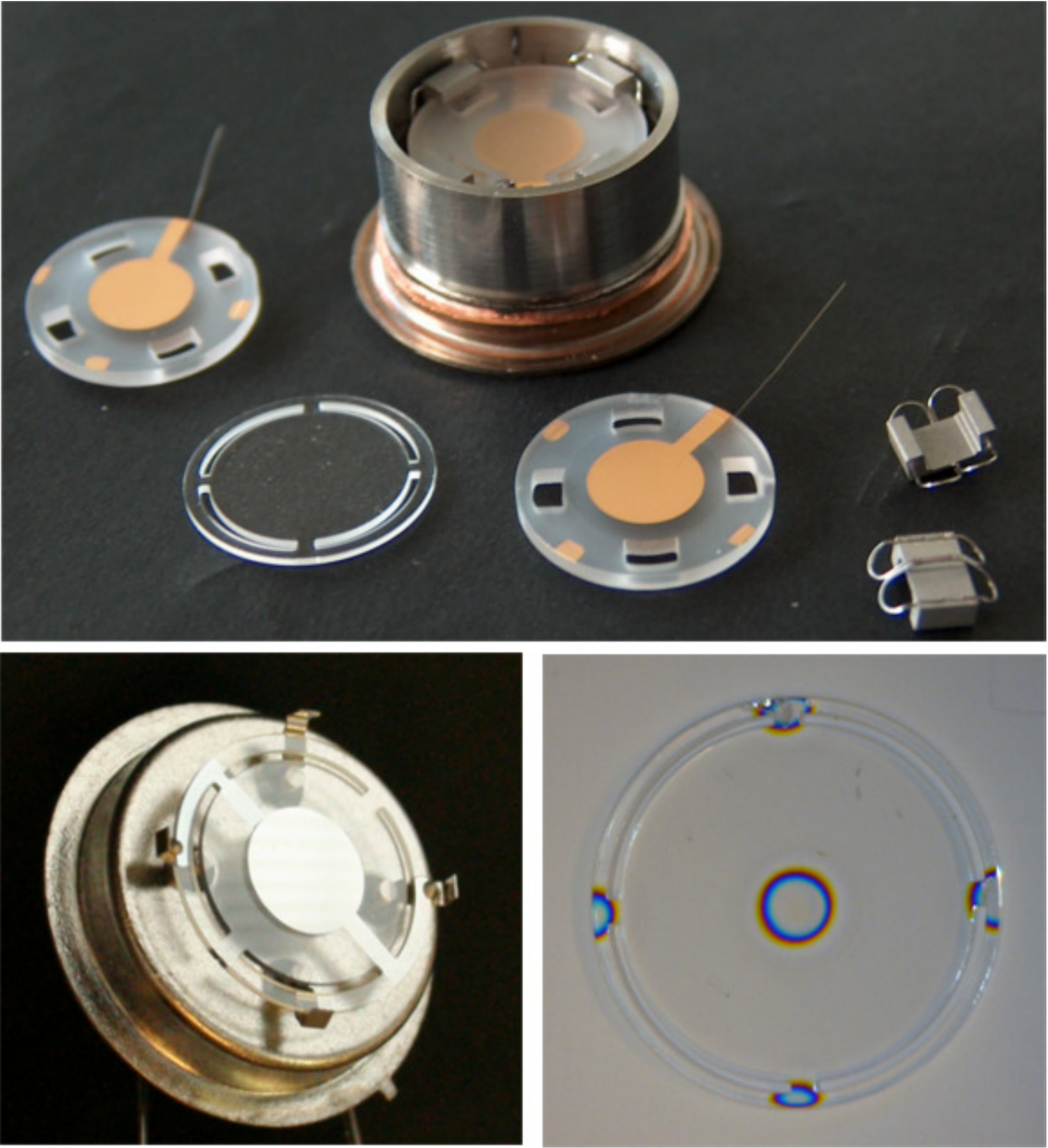}
\caption{\label{Devices_under_test_3}Examples of devices under test. Upper photograph: an electrodeless quartz resonator (whose outer diameter is 13.2 {\rm mm}).  Lower left photograph: a resonator with chromium electrodes; Lower right photograph: a resonator with ${\rm Ta}_2{\rm O}_5/{\rm Si}{\rm O}_2$ coating.}
\end{figure}

The tested devices are SC-cut quartz crystal resonators. They are 10 {\rm mm} diameter, 0.5 {\rm mm} tick, plano-convex disks made in a so-called ``premium-$Q$" synthetic crystalline quartz. Such a shape reduces losses in the resonator suspensions by trapping the energy of the generated bulk-acoustic-wave (BAW) within the disk centre~\cite{stevens1811} . They are optimized to resonate on the shear mode called C-mode, at 10 {\rm MHz}. Their $Q$-factor is typically $1.1\times10^6$ at 10 {\rm MHz} when operating under vacuum at room temperature. Once at 4 {\rm K}, their $Q$-factor can increase by two orders of magnitude or more, depending on several factors: material quality, surface roughness, stresses induced by the environment, etc~\cite{galliou091911, galliouScRip2013}.
Resonators have first been measured according to an electrodeless version, i.e. electrodes being maintained at less than two microns from both resonator surfaces by electrode quartz holders positioned on each side of the resonator. In a second step, both electrode holders have been removed and the resonator tested alone once coated by gold and then chromium for some of them and by a ${\rm Ta}_2{\rm O}_5/{\rm Si}{\rm O}_2$ coating for others (see Fig.~\ref{Devices_under_test_3}). Electrodes are deposited on two quartz holders. The latter are then set on each side of the resonator and maintained by means of small metallic clamps. Electrode holders are machined in order to maintain a gap of a few microns between the electrodes and the resonator surfaces.

\begin{figure}[t!]
\centering
\includegraphics[width=3.25in]{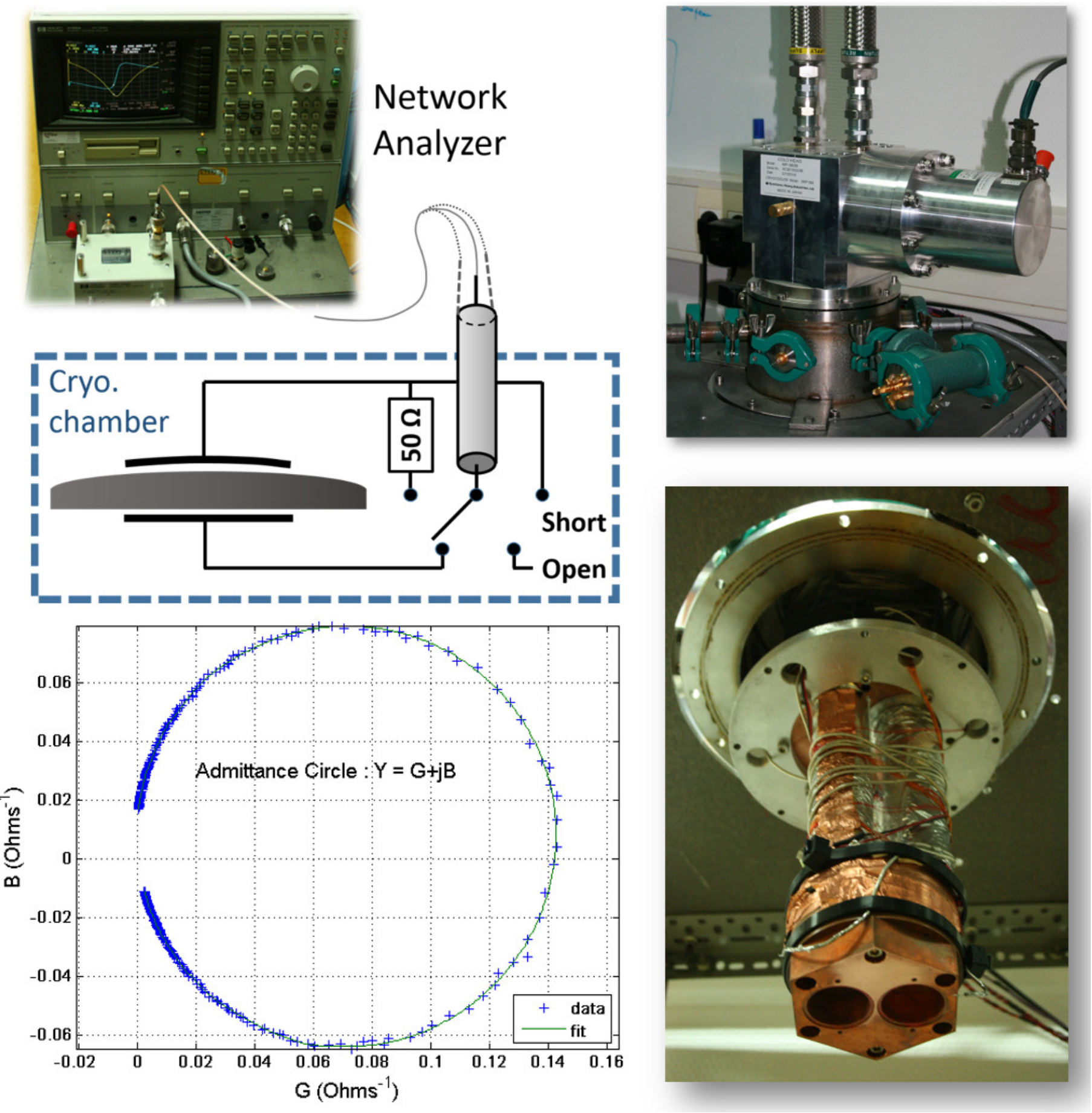}
\caption{\label{exp_setup4_gimp} Experimental set up and resulting plot after processing. Photographs show clockwise: the network analyzer; the outside part of the cryogenerator; the two stages of the cryogenerator once the vacuum enclosure removed, $G(B)$ data plot and a fit. }
\end{figure}

A conventional pulse-tube cryogenerator (SRP-082, SHI Cryogenics) is used to lower the temperature of the device under test down to the range $3.75$-$15$ {\rm K}, as illustrated in Fig.~\ref{exp_setup4_gimp}. The resonators under test are put inside an oxygen-free copper block whose temperature is controlled with $\pm 3$ {\rm mK} accuracy. The resonators and the copper block are in thermal equilibrium at each measurement point, and the pressure of their environment is of the order of $5\times10^{-7}$~{\rm mbar}. The resonators are connected to the measurement room temperate electronics using long coaxial cables thermalized at each stage of the cryocooler. Indeed, in the case of a quartz substrate, piezoelectricity offers a direct electrical readout of mechanical resonances. By means of a network analyzer a complex valued impedance is measured in the vicinity of different resonance frequencies at each temperature value. The technique incorporates a careful calibration stage that allows the method to exclude influence of the connecting cables. For this purpose three calibration standards are installed in the very proximity of the characterized device~\cite{ProcEFTF2008}. The technique employs a network analyzer HP4195A, and its impedance kit, locked to a Hydrogen Maser providing a short-term fractional frequency stability of $1\times10^{-13}$ over 1 {\rm s} combined with a long-term stability of $5\times10^{-16}$ over 10,000 {\rm s}. Special attention should be paid to the accuracy when characterising modes with extremely high quality factors due to extreme narrowness of their bandwidths. The frequency span of the analyzer is adjusted to get the maximum benefit of the 401 measurement points. The minimum available span of $0.5$ {\rm Hz} provides a resolution of about $1.25$~{\rm mHz}, and the sweep time as low as $10.5$ {\rm mn} is compatible with the expected unloaded $Q$-factors. 
Moreover, the driving power is kept as low as possible in order to avoid nonlinear and thermal effects. 
Power dissipation inside the resonator is estimated to vary from $0.1$ to $3$ {\rm nW}, depending on mode coupling and quality factor. The device characterization is performed using the impedance analysis method. In such a configuration the instrumentation provides $0.001$ {\rm dB}, and $0.01^\circ$ resolution in magnitude and phase respectively. Amplitude and phase data close to a resonance are then transformed in a $G(B)$ plot, i.e. the complex admittance $Y =G+jB$ is plotted in the complex plane. This provides a way to extract efficiently and accurately the corresponding quality factor in the case of a piezoelectric resonator~\cite{galliou091911, galliouScRip2013} (see Fig.~\ref{exp_setup4_gimp}).
Data are recorded from a network analyzer after a proper in-situ calibration procedure cancelling the effect of cables. The three cables ended by the open circuit, short circuit and 50 ohms load standards respectively are visible on the two stage photograph. The resulting admittance circle, i.e. the admittance $Y = G + jB$ plotted in the complex plane, is a fit of recorded points by using a neural network training method. Then, the $Q$-factor is calculated from the series frequency and the frequency bandwidth: the frequency corresponding to the maximum of the real part $G$ is the series resonance frequency $f_0$ and both cutting frequencies $f_1,f_2$ are extracted from the maximum and minimum of the imaginary part $B$, to get the Q value: $Q = f_0/(f_1-f_2)$.

\section{Results with metal coatings}

Losses in crystalline quartz depend on the material quality, i.e. the number of impurities and defects, and can vary from sample to sample. As a consequence, it is preferable to use the same substrates in various configurations instead of  comparing different resonators. After measuring $Q$-factors in an electrodeless configuration, i.e. with both electrodes being deposited on quartz holders and not on surfaces of the resonator itself, the usual Gold coating for electroded resonators have been performed on the resonator and $Q$-factors have been measured again. $Q$-factors have been measured each time for several OTs of the three resonance modes, the (quasi-)longitudinal one, namely the A-mode, and both (quasi-)shear modes, the B and C modes. The usual Gold coating implies two stage deposition: firstly, a 10 {\rm nm} layer of Chromium, that ensures a good adhesion of Gold, secondly, 140 {\rm nm} gold layer deposition. This gives a total coating thickness of 150 {\rm nm} on each surface of the resonator under test.
The same set of measurements has been performed after etching the Gold deposition on the resonators previously tested resonators. In the second run, Chromium has been coated to reach a total coating thickness of 35 {\rm nm} on both surfaces of the resonator.

\begin{figure}[t!]
\centering
\includegraphics[width=3.25in]{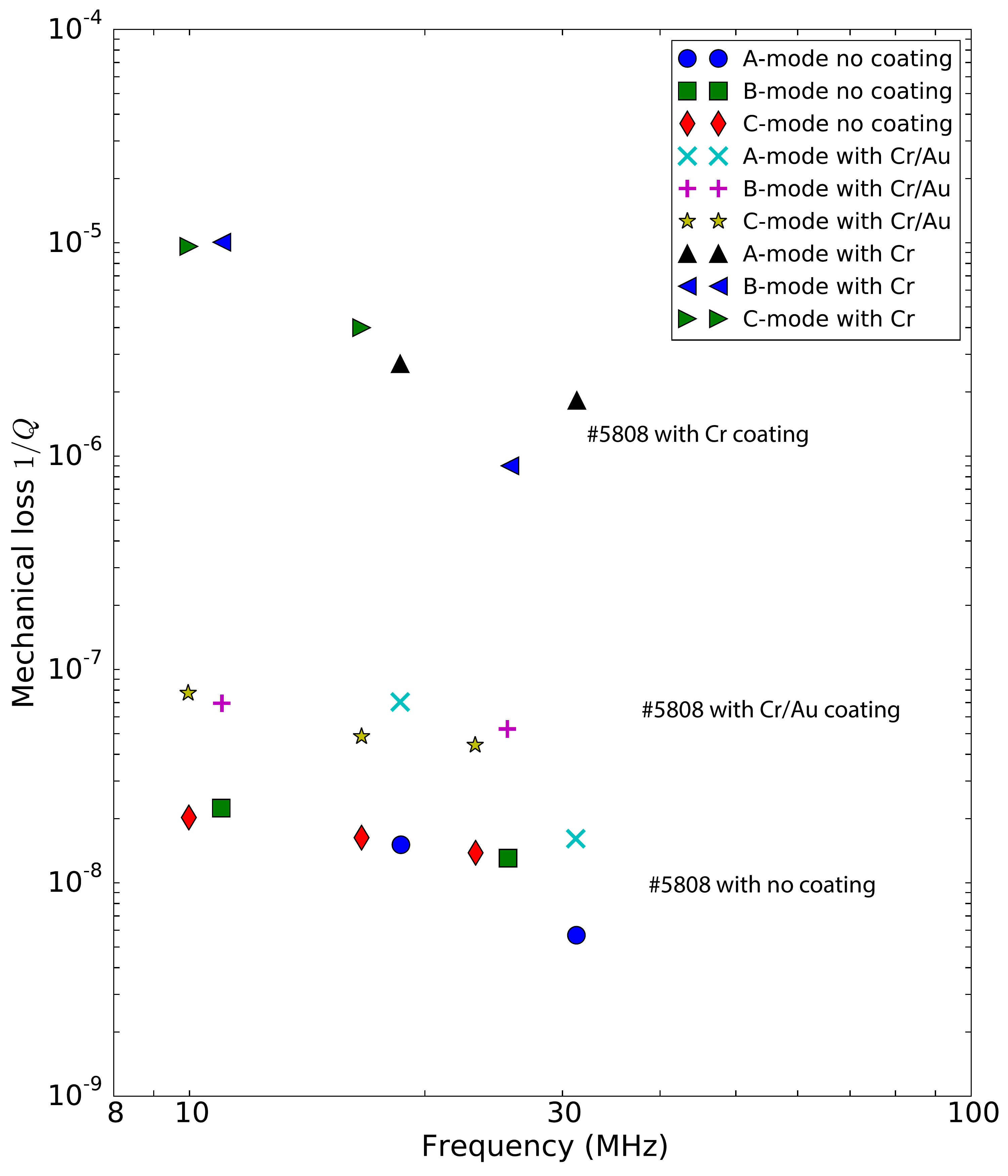}
\caption{\label{5808_loss_fev2016} Mechanical losses $\phi=1/Q$ versus frequency of low-order OTs of the A, B and C modes, at 4 {\rm K}, for sample $5808$ in three different situations: uncoated substrate, Gold evaporated on both surfaces, Chromium on both surfaces.}
\end{figure}

\begin{figure}[t!]
\centering
\includegraphics[width=3.25in]{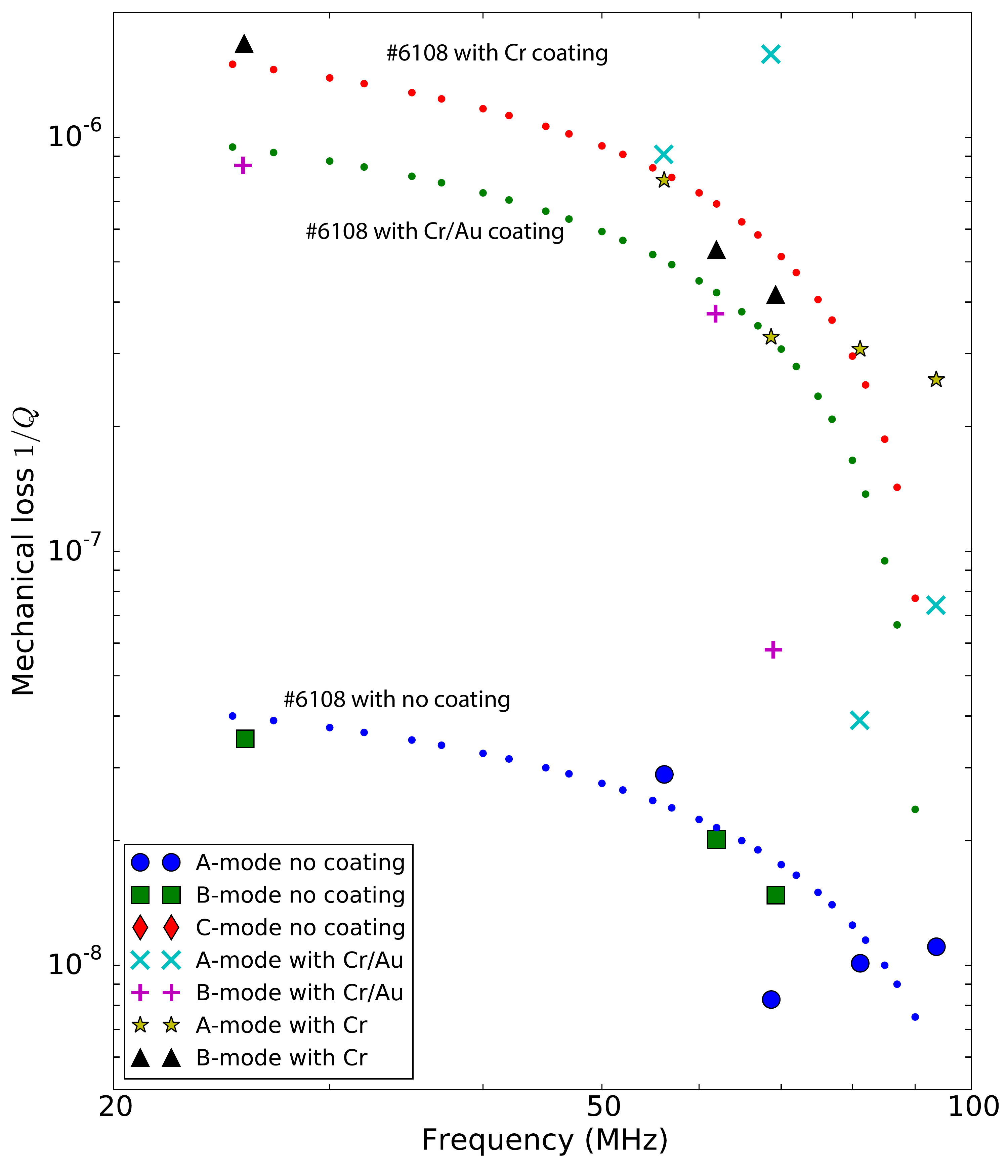}
\caption{\label{6801_loss_fev2016} Effect of gold and chromium coatings on sample $6801$ in comparison with the case of the uncoated substrate, in terms of mechanical loss $\phi=1/Q$ versus frequency, at 4 {\rm K}. Higher OTs than in Fig.~\ref{5808_loss_fev2016} are presented here for the A and B modes. Dotted lines are the linear regressions for the data. }
\end{figure}

 Results in terms of losses of various OTs for a given resonance mode before and after metal deposition for two different samples operating at 4 {\rm K} are shown in Fig.~\ref{5808_loss_fev2016} and Fig.~\ref{6801_loss_fev2016}. Different frequency ranges can be explored depending on the electrode diameter, a smaller electrode diameter leading to a smaller parasitic capacitor, and then enabling to explore resonances at higher frequencies (as an illustration, electrodes of sample $5808$, Fig.~\ref{5808_loss_fev2016}, are larger than those of sample $6801$, Fig.~\ref{6801_loss_fev2016})\cite{Goryachev:2013aa}. Beyond the overall trend of losses for a given configuration, which is outside the scope of this paper, most of the loss values clearly change in the same proportion from one configuration to the other, depending on the nature of the coating.

\section{Results with dielectric coatings}

The same methodology has been applied to compare the behavior of a quartz resonator with and without dielectric coating. A multilayer tantala/silica stack was coated only on one surface of the resonator without any postprocessing annealing. In both cases, external conducting electrodes deposited on both quartz holders were used to drive the resonator electrically. These electrodes were separated from the resonator surface by about $1 {\rm \mu m}$. The ${\rm Ta}_2{\rm O}_5/{\rm Si}{\rm O}_2$ coating of total thickness of 5615 {\rm nm} consists of 16 ${\rm \lambda/4}$ alternating layers of ${\rm Ta}_2{\rm O}_5$ and ${\rm Si}{\rm O}_2$. This coating has been done at Laboratoire des Mat\'{e}riaux Avanc\'{e}s, Villeurbanne, France, according to a well-established process.

\begin{figure}[t!]
\centering
\includegraphics[width=3.45in]{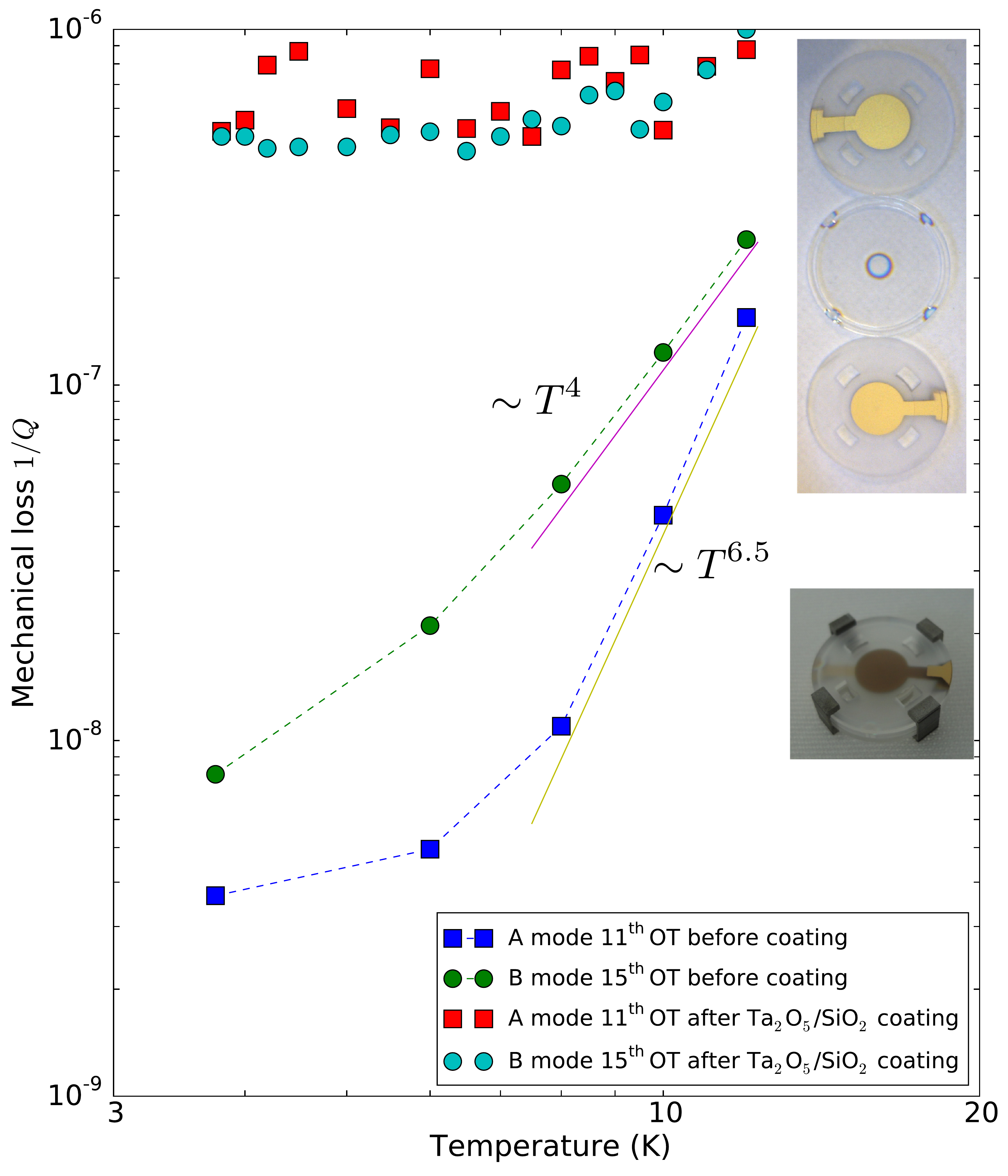}
\caption{\label{Dielectric} Mechanical losses $1/Q$ versus temperature of a quartz resonator without and with a single surface silica/tantala coating. OTs of A (extensional) and B (thickness shear) modes are at 34.244 {\rm MHz} and 27.280 {\rm MHz} respectively. The coated resonator is shown between both electrode holders in the left inset. }
\end{figure}

Fig.~\ref{Dielectric} shows the typical behavior of mechanical losses as a function of temperature for the A and B modes in the case of a bare and ${\rm Ta}_2{\rm O}_5/{\rm Si}{\rm O}_2$ coated quartz substrate. A ${\rm 1/T^n}$, where ${\rm 4<n<7}$, dependence for ${\rm 6\;K<T<12\;K}$ is the signature of three-phonon processes in the Landau-Ruomer regime~\cite{landaurumer1, Maris1971}. This power low demonstrates that the intrinsic dissipation in the bare quartz substrate are dominant and no suspension related losses are important. On the contrary, when coated with a dielectric multilayer, the resonator exhibits losses that are clearly limited by dielectric coating and do not have any clear temperature dependence. Fig.~\ref{Dielectric} demonstrates data for the $11^{\text{th}}$ OT of the A mode and the $15^{\text{th}}$ overtone of the B mode giving 34.244 {\rm MHz} and 27.280 {\rm MHz} resonance frequencies respectively. Note that the vibration energy is trapped at the centre of the resonator. The right inset shows the stack of the three quartz parts, ready to be tested once the electrical connections are done. It is important to underline that in the case of a dielectric-coated quartz substrate, mechanical losses have lower frequency dependence, i.e. the resonator coated results do not depend on the resonance mode. This fact justifies the use of the simplified model described in the next section.

\section{Discussion}

A simplified but efficient model has often been successfully used to assess the angle loss of a coating deposited on isotropic and even anisotropic substrates, mainly in order to determine the limitation due to thermal noise from ${\rm Ta}_2{\rm O}_5/{\rm Si}{\rm O}_2$ coatings on fused silica or sapphire, to provide mirrors for GW-detectors~\cite{HarryCQG2002, CrooksCQG2004, Pen2008, YamamotoPhysRevD2006}. It has been applied for data at room temperature~\cite{CrooksCQG2004, Pen2008}, as well as at low temperatures~\cite{YamamotoCQG2004, YamamotoPhysRevD2006, HirosePhysRevD2014}.
In these experiments, dependence of the mechanical loss of the dielectric coating with frequency has sometimes been considered~\cite{CrooksCQG2004}, but, anyway, the frequency range of interest never goes above a few kilohertz in those applications. It should be noticed that below a few kilohertz, internal friction (of a longitudinal or transverse wave) in an homogeneous medium is dominated by thermoelastic dissipation~\cite{Zener1938}. This is no longer the case when frequencies are greater than $5$ {\rm MHz} at low temperatures, like in the experiments discussed in this paper: phonon-phonon dissipation dominates inside the substrate~\cite{Maris1971, landaurumer1, galliouScRip2013, Deresiewicz1957, Zeller1971, Barron1982}.

Assuming that all the engineering losses are negligible and, in addition, that the interface damping is also negligible, the basic relationship is reminded as follows~\cite{Berry1981}:
\begin{equation}
\label{basic_equation}
\phi \approx \phi_{s}+\frac{E_{c}}{E_{s}}\phi_{c},
\end{equation}
where $\phi$, $\phi_{s}$, and $\phi_{c}$ are respectively the mechanical loss of the coated device, the quartz substrate (or uncoated device), and the coating itself, $E_{c}$ and $E_{s}$ are respectively energies stored in coating and the substrate whose ratio is typically small. Given the known ratio between these energies, one can estimate the intrinsic loss in coating materials from the measured loss values provided that the substrate losses are negligible. Comparing the results for bare and coated quartz resonators, this condition holds true for cryogenic conditions.

For isotropic materials with a Poisson's ratio much lower than unity, it can be shown that the ratio between the energies can be simplified reduced to $\frac{3 t_{c}Y_{c}}{t_{s}Y_{s}}$, where $Y_i$ denote the Young's modulus of element $i$, and $t_i$ its thickness~\cite{LandauLifshitz,YamamotoPhysRevD2006}. Quartz is only slightly anisotropic, so this simplified model remains relevant.
Furthermore, in the operating conditions, Young's modulus of a thin two-layer coating $Y_{c}$ can be simply calculated as an average of Young's modulus of both deposited materials, weighted by their respective thicknesses $t_1$ and $t_2$\cite{Pen2008, YamamotoPhysRevD2006}:
\begin{equation}
\label{Y_multilayer}
 Y_{c} \approx \frac{t_{1} Y_{1} + t_{2} Y_{2}}{t_{1} + t_{2}}.
\end{equation}
Thus, the Young's modulus of the gold coating with the small chromium layer can be evaluated as an average of the two. On the other hand, the values for the latter may vary a lot depending on the coating process and the resulting layer microstructure giving values that can be and different from those for bulk the material. Moreover, there is no sufficient data for these values for cryogenic temperatures.
The Young's modulus for Chromium thin films $Y_\text{Cr}$ varies from $90$ {\rm GPa}~\cite{Lintymer2003, Liang2007} to $240$ {\rm GPa}~\cite{Whiting1990, Liang2007} with $Y_\text{Cr}\approx 170$ {\rm GPa} being the most appropriated value in our case~\cite{Lintymer2003}. As for deposited Gold, $Y_\text{Au} \approx 75$ {\rm GPa} is a convergent value~\cite{Liang2007, Lu2001, Sandberg2005}. Given these values, the weighted average provides a Young's modulus of coating used in our experiments on both surfaces of the quartz substrates estimated by $Y_\text{CrAu} = 81$ {\rm GPa}.  
Similarly, in the case of the multi-layer dielectric coatings, The Young's modulus is estimated $Y_\text{DC} = 100$ {\rm GPa}, with $Y_{\text{Ta}_{2}\text{O}_{5}}=140$ {\rm GPa} and $Y_{\text{SiO}_2}=72$ {\rm GPa}~\cite{Martin1993, YamamotoPhysRevD2006, Abernathy2014}.
Finally, Young's modulus of the SC-cut quartz crystal substrate can be calculated from two the elastic coefficients $C_{ij}$ as $Y_{s}=C_{11}-\frac{2 C_{12}^{2}}{C_{11}+C_{12}}$ giving $Y_{s}=86.23$ {\rm GPa}.

\subsection{Mechanical loss of metal thin layers at 4 {\rm K}}

For both metal coatings, Eq.\eqref{basic_equation} should be applied only to one half of the resonator thickness, because of the coating is deposited on both substrate surfaces. 4 {\rm K} measurement results of the lower frequency resonator (Fig.~\ref{5808_loss_fev2016}) would provide mechanical losses of $\phi_\text{Cr}\approx 16\times10^{-4}$ for Chromium and $\phi_\text{Au}\approx 4\times10^{-4}$ for Chromium/Gold, but lead to outliers when taking into account the frequency dependence. Results from the second sample working at higher frequencies (Fig.~\ref{6801_loss_fev2016}) lead to more consistent results summarised as follows:
\begin{subequations} \label{Phi_Cr_and_Au}
\begin {align}
\phi_\text{Cr}~({\rm in}~10^{-6}) \approx -(26 \pm 9) \times f ({\rm MHz}) + (2410 \pm 380) \label{Phi_Cr}, \\
\phi_\text{Au}~({\rm in}~10^{-6}) \approx -(4 \pm 6)\times f ({\rm MHz}) + (515 \pm 303) \label{Phi_Au},
\end{align}
\end{subequations}
where $25~{\rm MHz} \leq f \leq 80~{\rm MHz}$, when the fractional uncertainty of Young's moduli is set at 10\% for values at 4 {\rm K} based on data measured at 300 {\rm K}.

\subsection{Mechanical loss of ${\rm Ta}_{2}{\rm O}_{5}$/${\rm Si}{\rm O}_{2}$ coating at 4-12 {\rm K}}

In the case of the multi-layer dielectric stack coated on one surface, the overall quartz thickness should be taken into account. According to the modelling and data described above, the dielectric coating loss can be estimated as $\phi_{c}\approx (1.6 \pm 0.3)\times10^{-5}$ with Young's modulus uncertainties of 10\%. Nevertheless, since only stacks of dielectric layers were deposited, the combined losses can be estimated as follows:
\begin{equation} \label{Phi_Ta_Si}
\ (0.58 \pm 0.17) \times \phi_{\text{Ta}_{2}\text{O}_{5}} + (0.42 \pm 0.12)\times \phi_{\text{SiO}_2} \approx (1.6 \pm 0.3)~10^{-5},
\end{equation}
without possibility to estimate the losses of individual material layers of ${\rm Ta}_{2}{\rm O}_{5}$ and ${\rm Si}{\rm O}_{2}$. 

According to Eq.\eqref{Phi_Ta_Si}, losses of ${\rm Si}{\rm O}_{2}$ are lower than $\phi_{\text{SiO}_2} < 0.4\times10^{-4}$, at 4-12 {\rm K}, for frequencies around 30 {\rm MHz}. This limit is distinct from $0.8\times10^{-4}$, a published value~\cite{Bartell1982,Topp1996} from measurements at 4-10 {\rm K} around 43 {\rm MHz}, but close enough to other reported value of $0.5\times10^{-4}$~\cite{CrooksCQG2004, Pen2008, Principe2015} obtained at room temperature and low frequencies. Other measurements give values of $5\times10^{-4}$ at 5 {\rm K} and 17-28 {\rm kHz}~\cite{ZimmerSchroter1976, Nawrodt1978, Franc2009} and  20 {\rm K}~\cite{HirosePhysRevD2014}) or $3\times10^{-4}$ at 14 {\rm K} and 10 {\rm kHz}~\cite{LMA2015}.

Similarly, relation $\phi_{\text{Ta}_{2}\text{O}_{5}}\approx 0.3\times10^{-4} - 0.72 \times \phi_{\text{SiO}_2}$ and Eq.\eqref{Phi_Ta_Si} provide the upper bound of $0.3\times10^{-4}$ for the losses in ${\rm Ta}_{2}{\rm O}_{5}$.
This value is much lower than data already reported, mainly close to a few $10^{-4}$, for room temperature~\cite{CrooksCQG2004, Pen2008, Franc2009, Principe2015, LMA2015} as well as for low temperature~\cite{YamamotoPhysRevD2006, Franc2009, MartinCQG2009}. Although tantala losses dominate in all these results, operating frequencies are typically lower than a few {\rm kHz} that is significantly lower than a few tens of {\rm MHz} as demonstrated in the present work.

\section{Conclusion}

Results of the present work on internal friction in Chromium and Chromium/Gold thin films at cryogenic temperatures for frequencies of a few tens of megahertz are an appropriate complement to data already available at lower frequencies and/or higher temperatures~\cite{GolovinHandbook}. Some discrepancies can exist because of various factors such as the annealing temperature or the coating method (thin films exhibiting mechanical properties different from those of bulk materials). Results reported in this paper (on the order of $\phi_{\text{Cr}}\approx 16\times10^{-4}$ and $\phi_{\text{Au}}\approx 4\times10^{-4}$) though slightly frequency dependent are still consistent with the partially existing previous data\cite{Uozumi1972, GolovinHandbook}. 
It should be mentioned that quality factors of metallized resonators used in frequency standards, i.e. high-quality devices in ultra-stable oscillators, differ from that for their uncoated counterparts on less than a few tenths of parts per million at room temperature. Although the effect of such metal coatings is quite subtle at room temperature, for cryogenic temperature operation, it is significantly enhanced, as demonstrated in this paper.

Regarding loss in tantala/silica coating, the single multilayer-coating exhibits loss of $\phi_{c}\approx 1.6\times10^{-5}$ at 4 {\rm K} and around 30 {\rm MHz}. On the other hand, such test does not provide individual losses for tantala and silica. Nevertheless, the measured coating loss exhibits a value more than one order of magnitude lower than those published at low frequency~\cite{YamamotoPhysRevD2006, Pen2008, MartinGWADW2014, HirosePhysRevD2014, Principe2015}.  

The presented method can be used for any type of coatings designed for operation at 10K and below. The distinguishing feature of the proposed method is that it probes higher frequency mechanical behaviour of thin films. The major limitation of the proposed method comes from the lack of knowledge of Young's moduli at cryogenic temperatures. 



\section*{Acknowledgements}

This work has been supported by R\'{e}gion de Franche Comt\'{e} (Grant No. 2008C16215), France, LNE (Laboratoire National de M\'{e}trologie et d'Essais), Paris, France (LNE/DRST 15 7 002), the Embassy of France, Canberra, Australia (Scientific Mobility Program 2014), and the Australian Research Council (grants CE110001013 and DP160100253). Dielectric coatings have been made by LMA (USR CNRS 3264), Lyon, France.


%

\end{document}